\documentclass{article}

\usepackage{arxiv}
\usepackage{authblk}
\usepackage[utf8]{inputenc} 
\usepackage[T1]{fontenc}    
\usepackage{hyperref}       
\usepackage{url}            
\usepackage{booktabs}       
\usepackage{amsfonts}       
\usepackage{nicefrac}       
\usepackage{microtype}      
\usepackage{lipsum}
\usepackage{mathtools}
\usepackage{amsmath,amssymb} 
\usepackage{amsthm}
\usepackage{graphicx}
\usepackage{mwe}
\title{Noise2Kernel: Adaptive Self-Supervised Blind Denoising using a Dilated Convolutional Kernel Architecture}
\author[1]{\textbf{Kanggeun~Lee}}
\author[2,*]{\textbf{Won-Ki~Jeong}}
\affil[1]{School of Electrical and Computer Engineering, UNIST, South Korea}

\affil[2]{Department of Computer Science and Engineering, Korea University, South Korea}

\affil[*]{wkjeong@korea.ac.kr, }

\begin{document}
\DeclarePairedDelimiter\ceil{\lceil}{\rceil}
\DeclarePairedDelimiter\floor{\lfloor}{\rfloor}

\theoremstyle{plain}
\newtheorem{thm}{Theorem}
\theoremstyle{definition}
\newtheorem{prop}[thm]{Proposition}
\newcommand{\QEDA}{\hfill\ensuremath{\blacksquare}}
\newtheorem{defn}{Definition} 

\maketitle

\begin{abstract}
 With the advent of recent advances in unsupervised learning, efficient training of a deep network for image denoising without pairs of noisy and clean images has become feasible. However, most current unsupervised denoising methods are built on the assumption of zero-mean noise under the signal-independent condition. This assumption causes blind denoising techniques to suffer brightness shifting problems on images that are greatly corrupted by extreme noise such as salt-and-pepper noise. Moreover, most blind denoising methods require a random masking scheme for training to ensure the invariance of the denoising process. In this paper, we propose a dilated convolutional network that satisfies an invariant property, allowing efficient kernel-based training without random masking. We also propose an adaptive self-supervision loss to circumvent the requirement of zero-mean constraint, which is specifically effective in removing salt-and-pepper or hybrid noise where a prior knowledge of noise statistics is not readily available. We demonstrate the efficacy of the proposed method by comparing it with state-of-the-art denoising methods using various examples.
\end{abstract}
\footnotetext{This work has been submitted to the IEEE for possible publication. Copyright may be transferred without notice, after which this version may no longer be accessible.}
\section{Introduction}

Denoising is one of the actively studied topics in computer vision and image processing. Images generated from various devices are prone to noise and corruption due to limited imaging environments (e.g., low light, slow shutter speed, etc). 
%
%
Conventional denoising methods usually rely on known noise models based on specific noise distributions. For instance, image prior based approaches, such as self-similarity~\cite{buades2005non, dabov2007,mairal2009non,dong2012nonlocally,gu2014weighted} require a specific property of pre-defined noise statistics or prior knowledge of a target image. However, there exist many real examples that pre-defined noise statistics do not fit, such as the coherent random noise observed in the transparent films used in Electron Microscopy (EM) imaging~\cite{minh2019removing}. 
In such cases, conventional denoising methods may not work well.

%


In recent years, the supervised learning of convolutional neural networks (CNNs) using clean-noisy image pairs has achieved superior denoising performance~\cite{zhang2017beyond,lefkimmiatis2018universal}. Due to the difficulty getting clean--noisy image pairs in real examples, a seminal work was completed by Lehtinen et al.~\cite{noise2noise} and introduced unsupervised learning of a denoiser (Noise2Noise (N2N)) using only noisy images.
%
%
Even though N2N proposed the general approach, it still suffers from the acquisition of noisy-noisy image pairs under known noise statistics. 

\begin{figure*}[t]
\includegraphics[width=16.5cm,keepaspectratio]{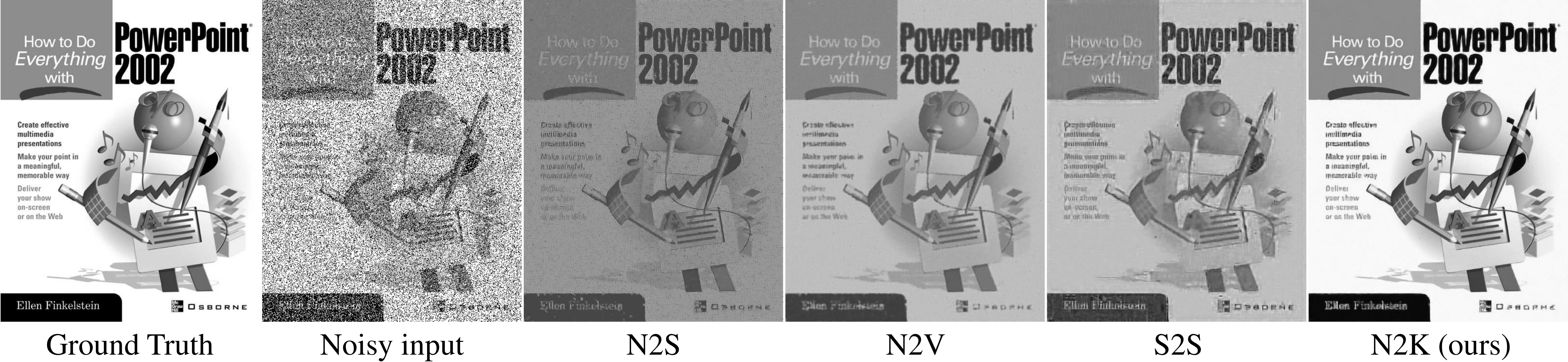}
\caption{Denoising results on the image highly corrupted by non-zero mean salt-and-pepper noise. Note that the results of N2S, N2V, and S2S look much darker than the ground truth. Our method (N2K) successfully removes noise without brightness shifting.}
   \label{fig:attract}
\end{figure*}

More recently, several new unsupervised image denoising methods~\cite{ulyanov2018deep,noise2void,noise2self,quan2020self2self} have shown promising results with denoisers that can be trained in a self-supervised fashion. 
For example, Noise2Void (N2V)~\cite{noise2void} and Noise2Self (N2S)~\cite{noise2self} require only the assumption of zero-mean noise without prior knowledge of noise statistics. These methods performed the denoising task successfully using only noisy images under a  zero-mean noise condition.
Self2Self (S2S)~\cite{quan2020self2self} proposed a novel framework with dropout using Bernoulli-sampled instances of a single input image. Moreover, using only a single training image, S2S outperformed on several famous noise distributions.
Despite their potential, these approaches have several drawbacks.
First, these self-supervised methods approximate the optimal denoiser with a noisy distribution based on the blind spot scheme (i.e., random masking of pixels during training). The blind spot scheme damages the original noisy image, and the large masking rate leads to poor performance.
%
%
%
%
Second is the weakness of the general self-supervision loss function. Because general self-supervision training depends on only the noisy signal, excessive noise causes CNNs to learn poorly and incorrectly. We discovered that state-of-the-art blind denoising methods are prone to predicting the wrong brightness or shape if noisy images are highly corrupted. Even though S2S successfully removed the pepper noise, it also predicted different brightness for salt-and-pepper noise in Fig.~\ref{fig:attract}. As a result, for state-of-the-art blind denoising methods, the brightness shifting artifact always appears in the case of highly corrupted non-zero mean noise (e.g., salt-and-pepper noise), as shown in Fig.~\ref{fig:attract}. 

To address the above issues, we introduce a novel unsupervised-learning-based denoising network, \texttt{Noise2Kernel} (N2K).
The combination of dilated convolution layers and donut-shaped kernels means it can build a specific function that satisfies the $\mathcal{J}$-$invariant$ property~\cite{noise2self}. 
%
In addition to the novel network architecture, we 
propose a novel adaptive self-supervision (ADSS) loss to restore the clean signal on the highly corrupted noisy image without brightness shifting.

The main contributions of our work can be summarized as follows:
\begin{enumerate}
\item We propose a novel dilated convolutional invariant 
network using a donut-shaped kernel and dilated convolutional layers. We no longer need a masking scheme and preprocessing to fill the masking region. 
\item We propose adaptive self-supervision loss, which is the pixel-level nonlinear energy, to suppress incorrect learning from noisy pixels. We demonstrate that the proposed adaptive loss is highly effective in preventing brightness shifting on images that are highly corrupted by noise, without any prior knowledge of the noise model. 
\item We demonstrate that our proposed method outperforms the state-of-the-art unsupervised denoising methods N2V, N2S, and S2S on various noise, such as salt-and-pepper, speckle, and the combination of several statistical noises (i.e., fusion noise).
\end{enumerate}
To the best of our knowledge, \texttt{Noise2Kernel} is the first fully blind denoising method that can prevent brightness-shifting for highly corrupted images without noise statistics and clean--noisy pairs.

\begin{figure*}[t]
\includegraphics[width=16.5cm,keepaspectratio]{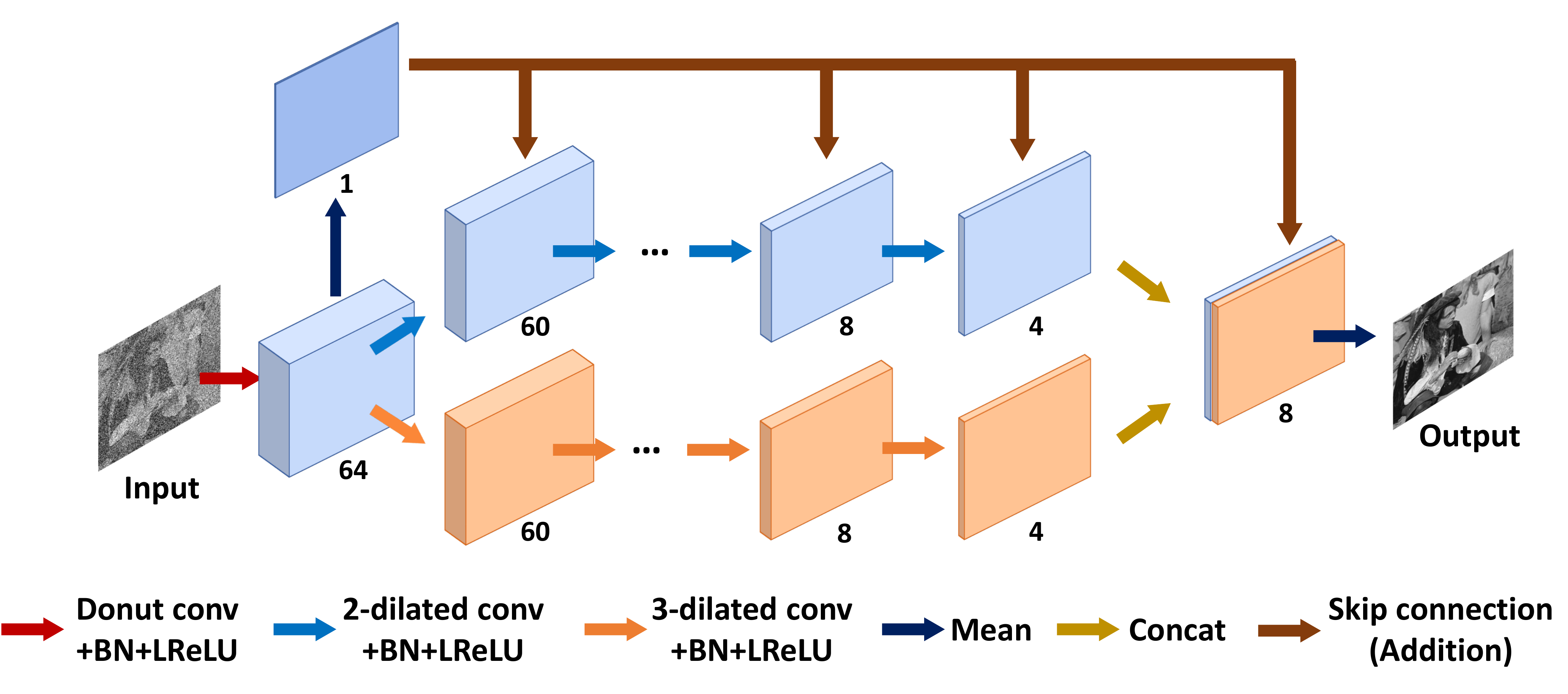}
\caption{Overview of the network structure of Noise2Kernel}
   \label{fig:structure}
\end{figure*}

\section{Related Work}
\subsection{Conventional Denoising Methods}
Total variation (TV), known as TV regularization, is a widely used denoising technique~\cite{vogel1996iterative,vese2003modeling,vese2004image,getreuer2012rudin} that adopts prior sparsity gradients in image denoising. 
Filtering methods~\cite{tomasi1998bilateral,buades2005non,dabov2007} based on spatial information or nonlocal self-similarity achieve better performance than TV-based methods. The block-matching and 3D filtering (BM3D) algorithm~\cite{dabov2007} still performs well enough to be used as a comparison for deep learning. The structure of BM3D is actively applied to various noise types, including salt-and-pepper noise and speckle noise~\cite{djurovic2016bm3d,parrilli2011nonlocal}. 
With enough training data, learning-based methods eventually perform better than non-learning-based ones. Before deep learning based on large training data, Dictionary Learning (DL) and Convolutional Sparse Coding (CSC)~\cite{wohlberg2017sporco} were used to restore the original signal using a sparse representation prior with a learned dictionary~\cite{bao2013fast, bao2015dictionary, elad2006image, papyan2017convolutional}.

\subsection{Non-blind Denoising Methods}
In recent years, with advances in deep learning and the related equipment, supervised deep learning over CNNs~\cite{zhang2017beyond,mao2016image,lefkimmiatis2017non} has shown great promise with its denoising performance. In contrast, it is not suitable to apply in practice because most supervised learning methods require noise statistics to generate the training data on a clean dataset.

Recently, the Noise2Noise (N2N)~\cite{noise2noise} proved that training a deep learning model is feasible and shows that the expected value of noisy inputs could be equal to the clean target.
However, in situations where the noise statistics are unknown, N2N is impractical because of the difficulty collecting a noisy pair for the same target.
With only noise statistics, these works~\cite{laine2019high,noisyasclean} perform as well as or slightly better than supervised learning. For instance, Laine \emph{et al.} ~\cite{laine2019high} suggest concrete self-supervision losses suitable for each noise statistic, but it is difficult to apply the proposed loss in cases with unknown noise statistics. 
Laine \emph{et al.} also presented a new blind spot network that makes a contribution similar to the architecture presented in this paper. 
However, we take a different approach to enabling self-supervision learning using the $\mathcal{J}$-$invariant$ property.
Similarly, Noisy-As-Clean (NAC)~\cite{noisyasclean} suggested a training scheme with pairs of noisy images $x$ and $x+n_s$ where $n_s$ is a simulated noise.
The researchers demonstrated that loss function $L(f(x+n_s),x)$ can be embedded into supervised learning.
Noisier2Noise~\cite{moran2020noisier2noise} presented a novel training approach with only a single noisy realization and noise statistics. It also overcomes the drawback of N2N that is the requirement of a prior of noise distribution. Moreover, the Noisier2Noise approach is applicable to spatially structured noise, one of the main disadvantages of a blind denoising method. 

\subsection{Blind Denoising Methods}
Blind denoising approaches assume that the prior noise distribution is unknown. 
To restore the clean signal without noise statistics, for instance, Deep Image Prior (DIP)~\cite{ulyanov2018deep} tries to utilize a handcrafted prior to the   image processing tasks. In other words, DIP shows that image prior can be learned by a random-initialized neural network without a specific condition. However, the internal image prior based approach has the two drawbacks of excessive testing time and inadequate performance.

The external image prior based approaches, such as N2V~\cite{noise2void} and N2S~\cite{noise2self}, employ the blind-spot scheme to prevent being an identity mapping function by self-supervised loss. 
Furthermore, two state-of-the-art methods take the self-supervision perspective to train the deep learning model using only noisy images. Two methods achieved significant shortening of testing time through the external image prior. In addition, N2S~\cite{noise2self} suggested the $\mathcal{J}$-$invariant$ property to prove that self-supervision loss can substitute for general loss of supervised learning. 

Recently, S2S~\cite{quan2020self2self} proposed a novel framework based on Bernoulli dropout, a new masking scheme in the training step, to avoid increasing the variance based on internal image prior because a single training sample causes large variances for denoising models such as a Bayes estimator.
Even though only a single noisy image is a training sample, S2S outperforms existing blind denoising methods based on the external image prior. 

\begin{figure*}[t]
\centering
\includegraphics[width=16.5cm,keepaspectratio]{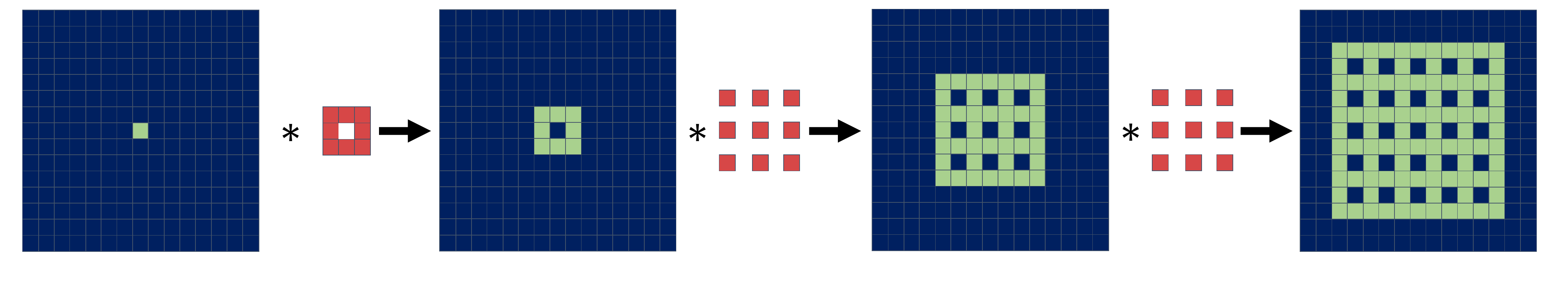}
\includegraphics[width=16.5cm,keepaspectratio]{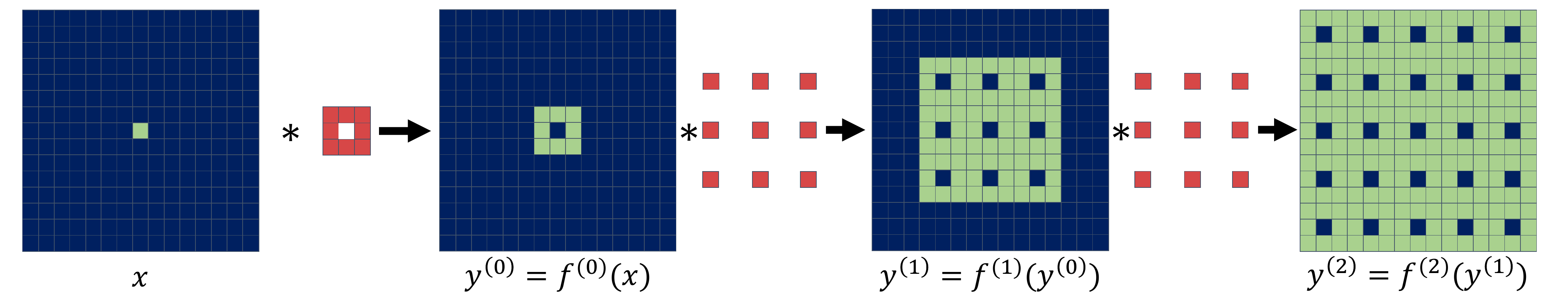}
{
 \caption{An example of dependency between the pixels in the input ($x$) and the output ($y$) images with one $3 \times 3$ donut convolution filter and $d$-dilated convolution. The each row represents the dependency visualization with two 2-dilated convolution and two 3-dilated convolution layers, respectively.
 The green pixels indicate the pixel locations having dependency with $x_{i,j}$ (the center pixel in $x$). %
 The red pixels represent the trainable variables of the convolution kernels. %
 The blue pixels indicates the area independent of $x_{i,j}$. 
 This figure shows the intermediate convolution processes of $y=f(x)=f^{(2)}(f^{(1)}(f^{(0)}(x)))$ from the input image to the output prediction. 
 (d) represents the final dependency of $x_{J}$ between the input and the output feature, showing no dependency with $y_{J}$.}
\label{fig:simulation}
}
\end{figure*}

\section{Method}

In this section, we introduce a novel deep neural network architecture that satisfies $\mathcal{J}$-$invariant$ property for blind noise reduction using adaptive self-supervision learning (Fig.~\ref{fig:structure}). 
First, we reiterate the definition of $\mathcal{J}$-$invariant$, originally introduced in  N2S~\cite{noise2self}. 
Next, we demonstrate that the proposed network satisfies the $\mathcal{J}$-$invariant$, which allows self-supervised training without using a specific training scheme (e.g., random masking).
%
Finally, we suggest the adaptive self-supervision loss to overcome the drawback 
of the conventional self-supervision loss on highly corrupted noisy images.

\subsection{Formulations}

This section introduces the formal definition and proposition of $\mathcal{J}$-$invariant$ that is required to explain the proposed network (more details can be found in~\cite{noise2self}). 
%
%
%
%
%
%
%
%
%


\begin{defn} 
Let $\mathcal{J}$ be a partition of the dimensions $\{1, ..., m\}$. 
Let $x$ be an observed noisy signal, and $x_J$ be a sub-sample of $x$ restricted to $J\in\mathcal{J}$. 
A function $f:\mathbb{R}^m\to\mathbb{R}^m$ is $J$-$invariant$ if 
the value of $f(x)_J$ does not depend on the value of $x_J$;
$f$ is $\mathcal{J}$-$invariant$ if it is $J$-$invariant$ for each $J\in\mathcal{J}$.
\end{defn}

We employ self-supervision loss as follows to restore the noisy image using the $\mathcal{J}$-$invariant$ function $f$. 
\begin{align}
    L(f) = \mathbb{E}||f(x)-x||^2
    \label{eq:selfloss}
\end{align}
To demonstrate that self-supervision loss can take the place of supervised loss, we borrow the same proposition from N2S under the $\mathcal{J}$-$invariant$ definition. 

\begin{prop} 
Let us assume that observed image $x$ is an unbiased estimator of $y$. 
Let $f$ be the $\mathcal{J}$-$invariant$ function. Then 
\begin{align}
    \mathbb{E}||f(x)-x||^2 = \mathbb{E}||f(x)-y||^2 + \mathbb{E}||x-y||^2
\end{align}

\end{prop}

\begin{proof}
Let us consider the self-supervision loss over $f$ function. 
\begin{align}
    \mathbb{E}_x||f(x)-x||^2 = \mathbb{E}_{x,y}||f(x)-y-(x-y)||^2 \nonumber \\ 
     = \mathbb{E}_{x,y}||f(x)-y||^2 + ||x-y||^2 - 2\langle f(x)-y,x-y \rangle \label{eq4}
\end{align}
The inner product term $\langle f(x)-y,x-y \rangle$ can be considered as follows:
\begin{align}
    \Sigma_{i} \mathbb{E}_y(\mathbb{E}_{x|y}[(f(x)_i-y_i)(x_i-y_i)]) \label{eq5}
\end{align}
Because $f(x)_i|y$ and $x_i|y$ are independent due to the invariant property of $f$,  Eq.~\ref{eq5} becomes $\Sigma_{i} \mathbb{E}_y(\mathbb{E}_{x|y}[f(x)_i-y_i])(\mathbb{E}_{x|y}[x_i-y_i])$. 
Then, the third term of Eq.~\ref{eq4} vanishes since $ \mathbb{E}_{x|y}[x_i-y_i]$ is zero due to the zero-mean assumption of noise. 
\end{proof}

From this, we can infer that the general self-supervised loss would be the sum of the general supervised loss and the variance of noise. 
Therefore, based on the similar scheme of~\cite{noise2self}, we can conclude that an invariant function $f$ can be a general denoiser if $f$ is minimized using a self-supervision loss. 
In the following section, we introduce the proposed network, which is an $\mathcal{J}$-$invariant$ function using a donut-shaped kernel based convolution layer and dilated convolutional layers.


\subsection{Dilated Convolutional $\mathcal{J}$-$invariant$ Network}
Assume that the function $f$ is a convolutional neural network with a single donut-shaped kernel (center weight is always zero) 
(see Fig.~\ref{fig:simulation}). 
%
Let $y$ be the unbiased estimator of the observed image $x$.
Based on Definition 1, the function $f$ satisfies the $\mathcal{J}$-$invariant$ property because $x_{i}$ is the sum of the multiplication of the neighboring information with the donut kernel ,except $x_{J}$, for all $J \in \mathcal{J}$ where the size of the squared donut kernel $K$ is always an odd number. 
We focus on this $\mathcal{J}$-$invariant$ function in a fully convolutional network (FCN).
If only one general convolution layer is added, the invariant property is not satisfied even though the first layer may use the donut kernel.
Furthermore, the receptive field of a single layer is too small to predict the correct pixel within the kernel. 
\begin{figure*}[t]
    \centering
    \begin{minipage}{.32\textwidth}
        \includegraphics[width=1\textwidth]{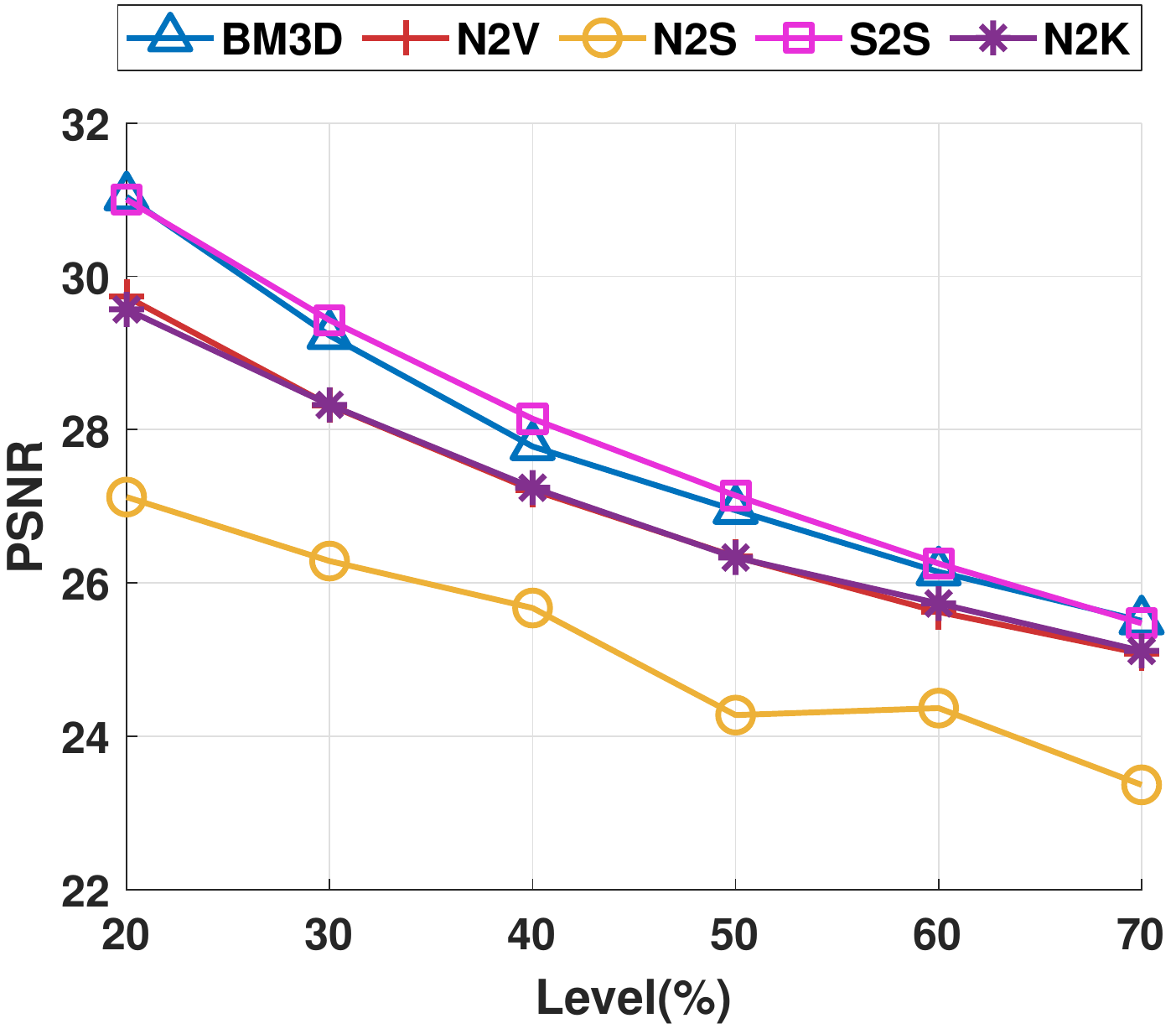}
    \end{minipage}
    \begin{minipage}{.32\textwidth}
        \includegraphics[width=1\textwidth]{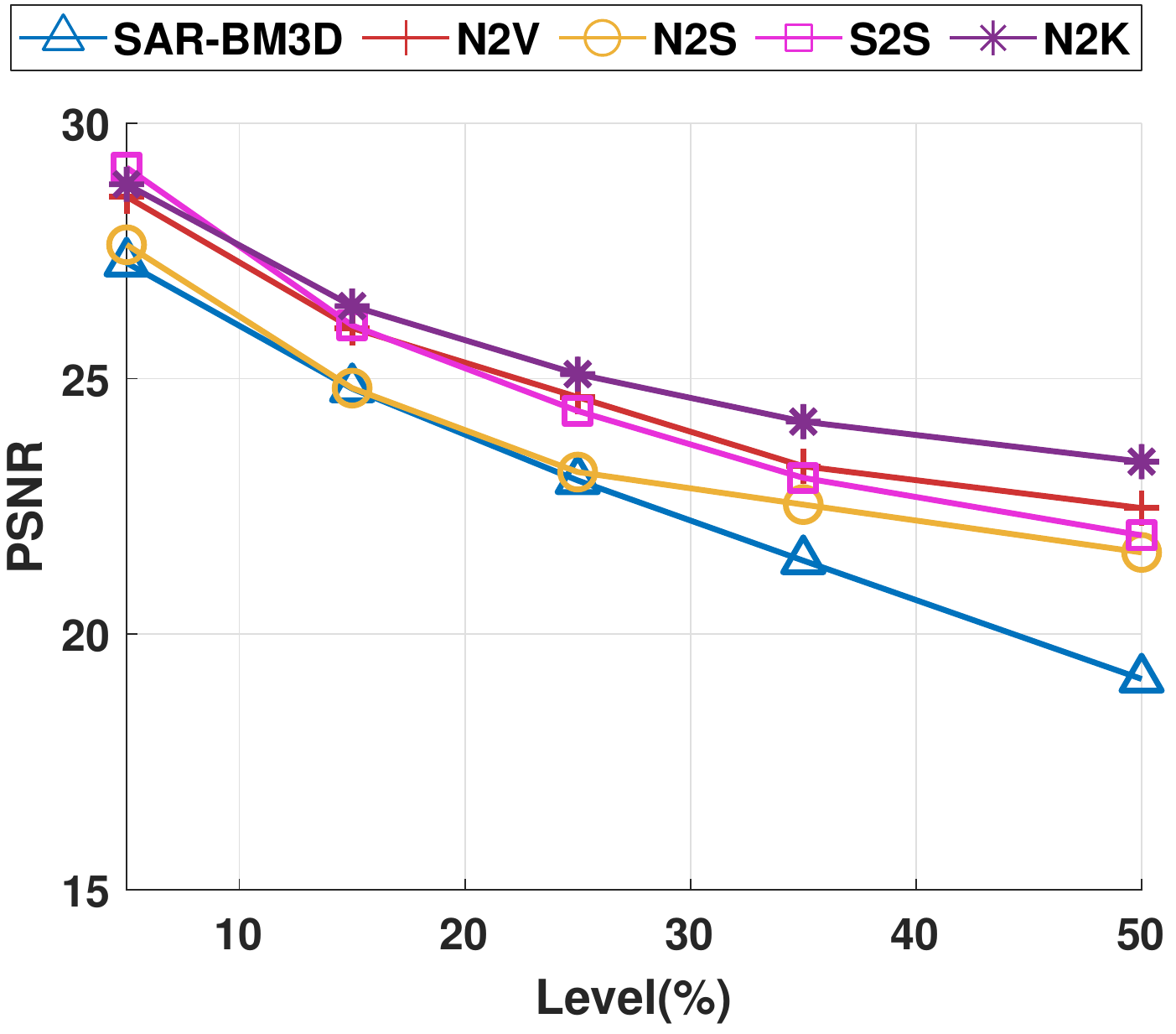}
    \end{minipage}
    \begin{minipage}{.32\textwidth}
        \includegraphics[width=1\textwidth]{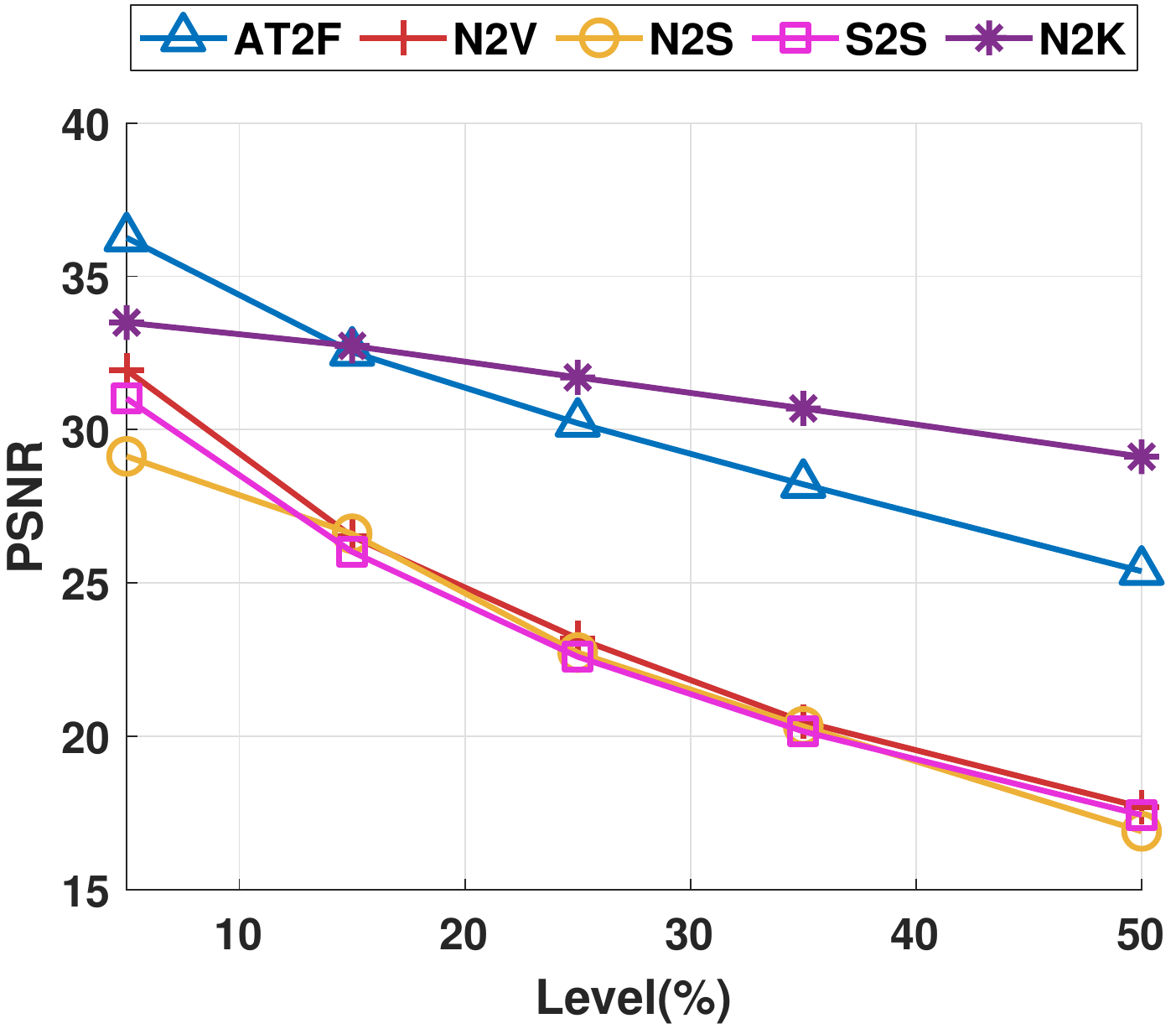}
    \end{minipage}
    \caption{Quantitative performance comparison of various denoising methods on known noise models in the Set14 dataset. Left to right: AWGN, speckle noise, and salt-and-pepper, respectively. }
    \label{fig:quantitative_result_known_noise_statistics}
\end{figure*}

Let $f$ be a network as a function that consists of $d$-dilated convolution $f^{(k)}$~\cite{yumulti} for all $k \in [1,n]$ where the size of the kernel is $3 \times 3$. 
We infer the function $f$ as $f(x)=f^{(n)}(f^{(n-1)}(...f^{(1)}(f^{(0)}(x))))$ where $f^{(0)}$ and $x$ are a convolution layer of the $K \times K$ donut-shaped kernel and an input noisy image, respectively, and $y^{(k)}$ is the output features for each $k$-th convolution layer.
We then need to demonstrate that $f(x)_{J}$ does not depend on $x_{J}$ for all $J \in \mathcal{J}$.

\begin{prop} 
The proposed network $f$ is $\mathcal{J}$-$invariant$ if $d \geq \ceil{K/2}$.
\end{prop}

\begin{proof} Without loss of generality, we consider a one-dimensional case to prove this proposition. 
Let us choose one pixel $x_J$ where $J \in \mathcal{J}$. Because of the donut convolution layer $f^{(0)}$, $x_J$ information moves to the neighboring region $\{J- \floor{K/2},...,J-1,J+1,...,J+\floor{K/2}\}$ as shown in Fig.~\ref{fig:simulation}. 
Let us suppose that the receptive field of $x_J$ in $y^{(k)}$ is $RF(y^{(k)},x_J)$. Then,
\begin{gather}
RF(y^{(k)},x_J) = \bigcup_{j\in\{-d,0,d\}} \{i+j|i\in RF(y^{(k-1)},x_J)\}
\label{eq:rec_relation}
\end{gather}
where $RF(y^{(0)},x_J)=\{J- \floor{K/2},...,J-1,J+1,...,J+\floor{K/2}\}$ for all $k\in[1,n]$.
By this recurrence relationship, we can infer that $\floor{K/2}-d<0$ and $-\floor{K/2}+d>0$ lead to exclude an element $J$ in $RF(y^{(n)},x_J)$. In other words, the $f(x)_{J}$ never consists of the information of $x_J$ if $d \geq \ceil{K/2}$.
\end{proof}
The combined structure of the donut convolution layer and dilated convolution layer, as shown in Fig.~\ref{fig:structure}, always guarantees the existence of the $\mathcal{J}$-$invariant$ property if $d \geq \ceil{K/2}$ and the size of the square kernel of donut convolution layer $K$ is an odd number.
In addition, as shown in Fig.~\ref{fig:structure}, there are two paths that both consist of 2-dilated or 3-dilated convolutional layers only. 
Because each path satisfies the $\mathcal{J}$-$invariant$ condition, the proposed network is $\mathcal{J}$-$invariant$. According to Eq.~\ref{eq:rec_relation}, the receptive fields of 2-dilation and 3-dilation paths are complementary for each pixel. 
%
%
To preserve the first prediction computed by a donut kernel, we added a skip connection after the dilated convolution operation. 
We discovered that the skip improved the convergence speed and image quality.
In addition to 
its model architecture, another important benefit of N2K is that it no longer requires the masking scheme. 
A masked input $\widetilde{x}$ of the noisy image $x$ with the dimension $J\subset \mathcal{J}$ (chosen randomly) is defined as
\begin{align}
    \widetilde{x} = \left\{ \begin{array}{rcl} 
 0 & \mbox{for} & j\in J  \\
 x_j & \mbox{for} & j\notin J 
\end{array}\right.
\end{align}
Then, the general self-supervision loss with the masking scheme is defined as follows:

\begin{align}
    \min_{\theta}\sum_{i}^{N}\sum_{J\subset\mathcal{J}}||(f_J(\widetilde{x}^i;\theta)- x^i_J)||^2
\end{align}

Because random pixel discarding in the masking scheme introduces defects in image ($\widetilde{x}$), N2V and N2S fill in these missing pixels by copying from random locations or through interpolation from neighboring pixels.
%
%
%
Unlike such existing methods, the dilation convolution architecture in N2K can be trained using only the original $x$ and the general self-supervision loss without a masking scheme shown below: 
\begin{align}
    \mathcal{L}(f) = \mathbb{E}||f(x)-x||^2 
\end{align}

\begin{figure*}[t]
\includegraphics[width=16.5cm,keepaspectratio]{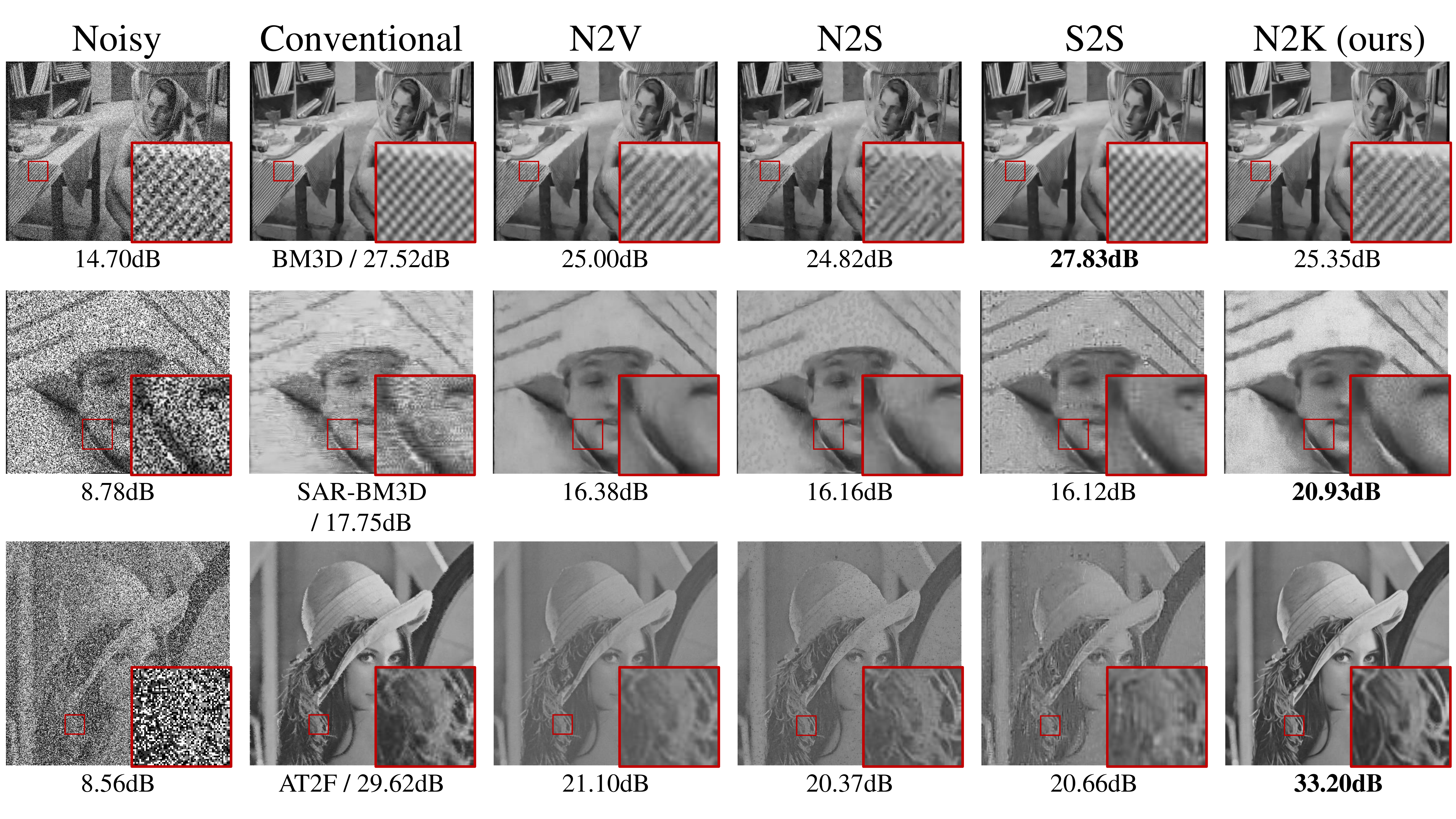}
\caption{Qualitative performance comparison of various denoising methods on three noise types. Top to bottom row: AWGN($\sigma_g$=50), speckle noise ($\sigma_s$=50), and salt-and-pepper noise ($d$=50), respectively. The best PSNR in each case is highlighted in bold.
Each row indicates the results of AWGN, speckle noise, and salt-and-pepper noise, respectively.}
  \label{fig:qualitative_for_three}
\end{figure*}

\begin{figure*}
\includegraphics[width=16.5cm,keepaspectratio]{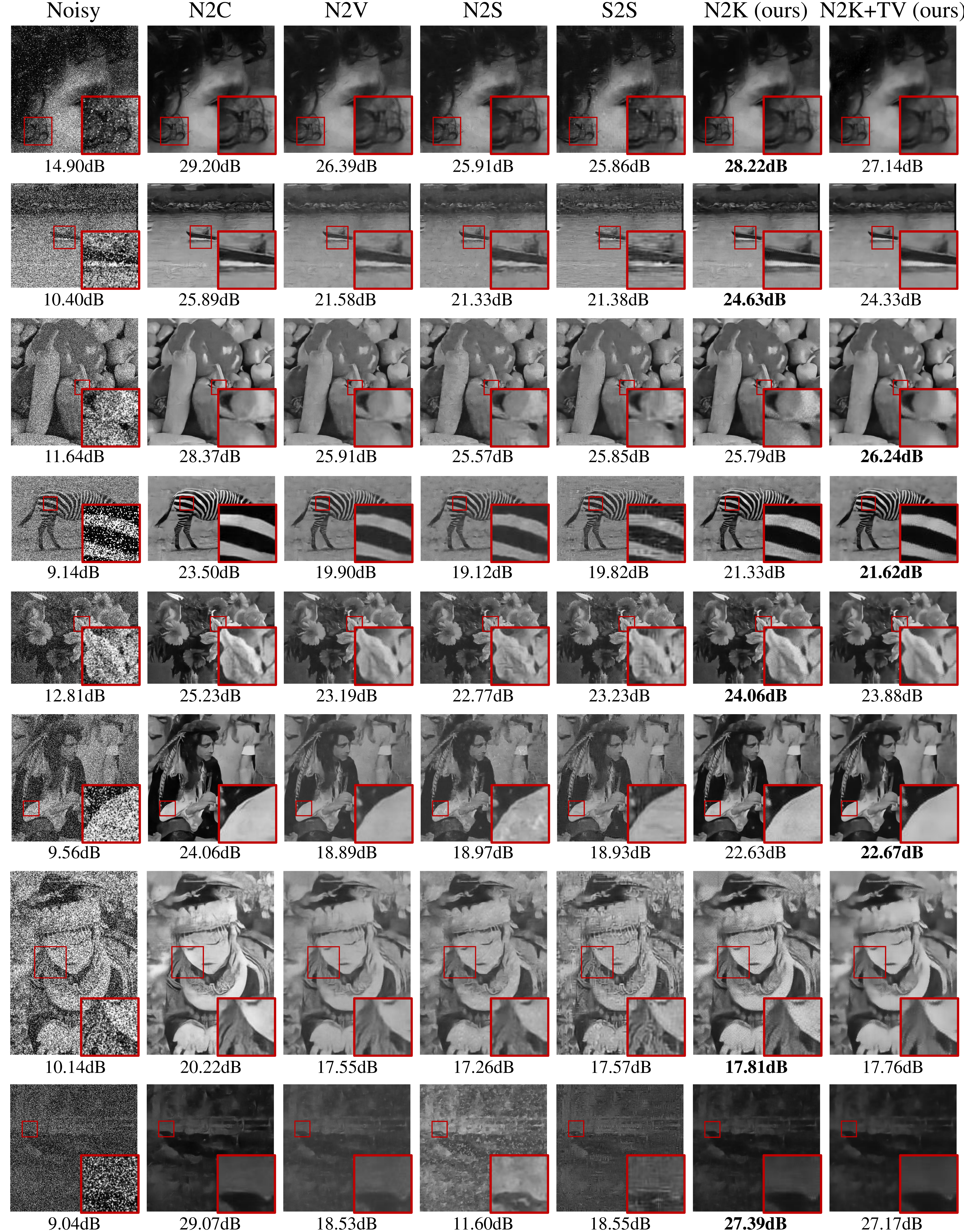}
\caption{Qualitative performance comparison of various denoising methods on fusion noise. Top to bottom ($\sigma_g$, $\sigma_s$, d): (25,5,5), (25,5,25), (25,25,5), (25,25,25), (50, 5, 5), (50, 5, 25), (50, 25, 5), (50, 25, 25), respectively. The best PSNR is highlighted in bold, except N2C.}
  \label{fig:qualitative_for_fusion}
\end{figure*}

\subsection{Adaptive Self-Supervision Loss}

In the unsupervised denoising problem, the zero mean noise is considered a default noise model. 
%
However, the zero mean condition is too strict to use on blind denoising with self-supervision loss.
%
For example, in the case of salt 
noise (i.e., random white dots), the general self-supervision loss may falsely treat the correct prediction as a noisy label due to large differences between the predicted and noisy pixel values, which causes brightness shifting toward white. 
%
%
%
This implies that self-supervision may fail to work on highly corrupted non-zero mean noise. 
To address such limitations of standard self-supervision loss, we propose ADSS loss 
using the focusing parameter $\lambda$ as follows:
%
\begin{gather}
    \mathcal{L}(f,w) = \frac{1}{N}\sum_{i}^{N}\sum_{J}w^{i}_{J}(f_J(x^{i})-x^i_J)^2 \\
    w^{i}_J = \dfrac{1}{1+\lambda|f_J(x^{i})-x^{i}_J|}
    \label{eqation:adss_weight}
\end{gather}

The ADSS loss adjusts the proportion of difference between $x$ and $f(x)$ adaptively.
The main idea behind ADSS is that, if the prediction is significantly different from the input pixel value, it is highly likely that the input pixel is noise. 
Therefore, during the training process, backpropagation from such pixels should be suppressed (i.e., the correct predictions should not be shifted toward the noise pixel values) by adaptive control of the weight in the loss function.
The ADSS loss is equivalent to the self-supervision loss when $\lambda = 0$. 
Intuitively, $\lambda$ controls the extent of the influence of discrimination. 



\section{Results}

To assess the performance of the proposed method, we tested on various noise models, such as additive white Gaussian noise (AWGN), speckle noise, and salt-and-pepper noise. 
In particular, because we focus on highly corrupted noisy images in the blind aspect, the noise should be modeled by unknown distribution. 
To simulate this, we built a fusion noise model by mixing AWGN, speckle noise, and salt-and-pepper noise. 
We compared our proposed method (N2K) with several state-of-the-art blind denoising methods (N2V, N2S, and S2S). 
In addition, we also compared N2K with conventional denoising methods, such as BM3D~\cite{dabov2007}, SAR-BM3D~\cite{parrilli2011nonlocal}, and AT2F~\cite{singh2018adaptive}, known for the best performing filter-based denoising method specifically designed for each noise model. 
%
%
%
We implemented Noise2Clean (N2C) on our own using the same network structure as shown in Fig.~\ref{fig:structure}, with a regular $3 \times 3$ convolution kernel for supervised training using the clean--noisy pairs introduced in Sec.~\ref{sec:blind_noise}. 
Note that N2C is a supervised-learning method, which serves as the upper bound for the performance of the learning-based denoising method.

For all training (except for N2C), we used only noisy images corrupted by simulated noise. We chose the same dataset, BSD400, of gray scale images used in~\cite{zhang2017beyond} and~\cite{noise2void} as a training dataset. For more detail, we applied augmentation using rotation and mirroring for all learning-based methods. For all testing of performance, we employed the Set14 dataset. We utilized the BSD68 dataset for the ablation study as a validation set.

We used TensorFlow~\cite{abadi2016tensorflow} (version 2.0.0) to implement the proposed architecture of N2K, as shown in Fig.~\ref{fig:structure}. 
For stable performance, we applied an exponential learning rate decay with an RAdam~\cite{liu2019variance} optimizer. 
%
We used batch size 64 and 3e-2 as the initial learning rate and 
$\lambda=10$ for Eq.~\ref{eqation:adss_weight}. 
For a fair comparison, we used the default parameter settings from the authors' code for other blind denoising methods. 
We picked the best hyper-parameters for experimental comparison methods when the setting of a hyper-parameter was required.
Because the denoiser should satisfy rotation invariance, we rotated each test image by 90 degrees and made two mirrored versions. 
The average of the inverse of eight outputs was the final prediction.

\subsection{Denoising Results on Known Noise Models}
\noindent
\subsubsection{Additive white Gaussian noise (AWGN)} 
AWGN is a popular statistical noise model with a zero mean characteristic as follows: 
\begin{align}
y = x+n,~n \sim \mathcal{N}(0,\,\sigma^{2}_{g})\,
\label{eq:gaussian}
\end{align}
where $\mathcal{N}$ is a normal distribution with standard deviation $\sigma_g$.
For the baseline performance, we chose BM3D, which is known for the best performing method for this noise model. 
%
%
For a fair comparison, we used the standard deviation $\sigma_g$ (i.e., noise level) of the given noise-corrupted images only for the case of BM3D (without the  noise level prior, BM3D does not produce correct results).
%
Fig.~\ref{fig:quantitative_result_known_noise_statistics} shows the quantitative performance comparison of denoisers over various noise levels, $\sigma_g$ from 20 to 70. 
N2K achieves similar or better performance than N2V and N2S, which are the state-of-the-art blind denoising methods, while S2S and BM3D outperformed N2K.
Therefore, we conclude that under the zero mean noise constraint, our method is comparable to most of the blind denoising methods except S2S. 
%
%
%
%

\begin{table*}[!t]
\centering
\renewcommand{\arraystretch}{1.3}
\renewcommand{\tabcolsep}{9.5pt}
\caption{Performance of baselines, N2K and N2K+TV on the Set14 test set. Boldface denotes the best among all except N2C.}
\label{table:fusion}
\centering
\begin{tabular}{||c|c|c|c|c|c|c|c|c||}
\hline
Noise level & \multicolumn{2}{c|}{$\sigma_g$=25, $\sigma_s$=5, $d$=5} & \multicolumn{2}{c|}{$\sigma_g$=25, $\sigma_s$=5, $d$=25} & \multicolumn{2}{c|}{$\sigma_g$=25, $\sigma_s$=25, $d$=5} &
\multicolumn{2}{c||}{$\sigma_g$=25, $\sigma_s$=25, $d$=25}\\ \hline
Method\textbackslash Metric & PSNR    & SSIM    & PSNR    & SSIM    & PSNR    & SSIM    & PSNR    & SSIM        \\ \hline \hline
N2C            & 28.06   & 0.7749   & 27.23  & 0.7476   & 25.61   & 0.6918   &  24.79  & 0.6615       \\ \hline \hline
N2V~\cite{noise2void}            & 25.51   & 0.7074  & 20.93  & 0.6089  & 22.59   & 0.6199  & 19.32   & 0.5335       \\ \hline
N2S~\cite{noise2self}            & 24.06   & 0.6683 & 20.40  & 0.5805  & 21.34   & 0.5797  & 18.68   & 0.4971       \\ \hline 
S2S~\cite{quan2020self2self} & 25.72 & \textbf{0.7256} & 20.88 & 0.5951 & 22.58 & 0.6252 & 19.27 & 0.5149 \\ \hline \hline 
N2K                              & \textbf{26.42}   & 0.7169  & \textbf{25.46}  & 0.6674  & 23.25   & 0.5782  & 22.19   & 0.4992       \\ \hline
N2K+TV                           & 26.26   & 0.7163  & 25.33  & \textbf{0.6791}  & \textbf{23.52}   & \textbf{0.6372}  & \textbf{22.67}   & \textbf{0.5966}       \\ \hline
\end{tabular}

\vspace{0.1in}

\begin{tabular}{||c|c|c|c|c|c|c|c|c||}
\hline
Noise level & \multicolumn{2}{c|}{$\sigma_g$=50, $\sigma_s$=5, $d$=5} & \multicolumn{2}{c|}{$\sigma_g$=50, $\sigma_s$=5, $d$=25} & \multicolumn{2}{c|}{$\sigma_g$=50, $\sigma_s$=25, $d$=5} &
\multicolumn{2}{c||}{$\sigma_g$=50, $\sigma_s$=25, $d$=25} \\ \hline
Method\textbackslash Metric & PSNR    & SSIM    & PSNR    & SSIM    & PSNR    & SSIM    & PSNR    & SSIM        \\ \hline \hline
N2C            & 26.16   & 0.7029   & 25.19  & 0.6631   & 24.66   & 0.6497   &  23.62  & 0.6082       \\ \hline \hline
N2V~\cite{noise2void}            & 23.01   & 0.6192  & 19.67  & 0.5572  & 20.88   & 0.5699  & 18.31   & 0.5005       \\ \hline
N2S~\cite{noise2self}            & 21.65   & 0.5714 & 19.07  & 0.4994  & 19.26   & 0.5103  & 17.76   & 0.4438       \\ \hline 
S2S~\cite{quan2020self2self} & 23.31 & 0.6441 & 19.64 & 0.5361 & 21.04 & 0.5725 & 18.35 & 0.4748 \\ \hline \hline 
N2K                              & \textbf{24.24}   & 0.6305  & 23.22  & 0.5708  & 21.20   & 0.5248  & 20.38   & 0.4438       \\ \hline
N2K+TV                           & \textbf{24.24}   & \textbf{0.6471}  & \textbf{23.35}  & \textbf{0.6076}  & \textbf{21.31}   & \textbf{0.5944}  & \textbf{20.49}   & \textbf{0.5452}       \\ \hline
\end{tabular}
\end{table*}

\subsubsection{Speckle noise} 
Signal-dependent multiplicative speckle noise, often observed in synthetic aperture radar and ultrasound images, can be modeled as follows: 
\begin{align}
y = x+n*x,~n \sim \mathcal{U}(0,\,\sigma^{2}_{s})\, \label{eq:speckle}
\end{align}
where $\mathcal{U}$ is the uniform distribution with a zero mean and a standard deviation of $\sigma_s$. 
We chose SAR-BM3D~\cite{parrilli2011nonlocal}, one of the conventional denoising methods specifically designed for speckle noise, as the baseline method to compare with N2K. 
%
%
We conducted the denoising experiment over various noise levels  $\sigma_s$ from 5 to 50. 
Interestingly, blind denoising methods outperformed SAR-BM3D, as shown in the second column 
of Fig.~\ref{fig:quantitative_result_known_noise_statistics}. 
%
Note that N2K consistently outperformed the other blind denoising methods for all noise levels $\sigma_s$ we tested (see the middle graph in  Fig.~\ref{fig:quantitative_result_known_noise_statistics}).
%
Furthermore, the 
performance gap 
between blind denoising methods and SAR-BM3D increases as the noise level increases, which implies that blind denoising methods are more robust to strong speckle noise than SAR-BM3D. 
%
N2K achieved the best difference 
compared to other blind denoisers (by around 4.55 dB higher) on the foreman image in the Set14 dataset (the second row of Fig.~\ref{fig:qualitative_for_three}). 
Moreover, the overall intensity distribution in 
the predicted image of N2K is closer to that of the ground truth; those of other blind denoisers (N2V, N2S, and S2S) suffer from brightness shifting due to the non-zero mean noise characteristic. 
%


\subsubsection{Salt-and-pepper noise} In this experiment, we employed salt-and-pepper noise, defined as follows:
\begin{align}
y = f_{spn}(x,d) \label{eq:salt_pepper}
\end{align}
where $f_{spn}$ is the projection function set to 0 or 1 with probability $d$. 
Conventional nonlinear denoising methods for salt-and-pepper noise, such as median filter or AT2F, work well on this noise model. 
We conducted the experiment using various noise levels from 5\% to 50\%. 
For the salt-and-pepper noise, N2K performed better than state-of-the-art methods because of its ability to overcome the problem of brightness shifting, as shown in Fig.~\ref{fig:qualitative_for_three} (third row).
Note that other blind denoising methods (N2V, N2S, and S2S) performed poorly on this noise model. 
Furthermore, N2K outperformed AT2F when $d\geq15$ on Set14, 
as shown in Fig.~\ref{fig:quantitative_result_known_noise_statistics}. 
Similar to speckle noise, blind denoising methods (i.e., N2V, N2S, and S2S) failed to restore the image contrast of the clean image but N2K successfully preserved the contrast and brightness of the original image. 
Note also that the results of AT2F 
look much blurrier than those of N2K.


\begin{table*}[t]
\centering
\renewcommand{\tabcolsep}{14pt}
\renewcommand{\arraystretch}{1.3}
\caption{Comparison of ADSS and general self-supervision loss. Average PSNR and SSIM for fusion noise on BSD68 validation set. The baseline uses only the structure of N2K with general self-supervision $L_2$ loss.}
\label{table:ablation_adss}
\begin{tabular}{||c|c|c|c|c|c|c||}
\hline
Model & \multicolumn{2}{c|}{Baseline} & \multicolumn{2}{c|}{N2K} & \multicolumn{2}{c|}{N2K+TV} \\ \hline
Noise level\textbackslash Metric 
& PSNR    & SSIM    & PSNR    & SSIM    & PSNR    & SSIM \\ \hline \hline
$\sigma_g$=25,$\sigma_s$=5,$d$=5   & 24.54   & 0.6761   & \textbf{25.28}   & \textbf{0.6892}   & 25.13   & 0.6853 \\ \hline 
$\sigma_g$=25,$\sigma_s$=5,$d$=25   & 20.93   & 0.5577   & \textbf{24.52}   & 0.6435   & 24.42   & \textbf{0.6513} \\ \hline 
$\sigma_g$=25,$\sigma_s$=25,$d$=5   & 21.66   & 0.5679   & \textbf{22.42}   & 0.5580   & 21.61   & \textbf{0.6043} \\ \hline 
$\sigma_g$=25,$\sigma_s$=25,$d$=25   & 19.22   & 0.4850   & 21.46   & 0.4869   & \textbf{21.86}   & \textbf{0.5673} \\ \hline 
$\sigma_g$=50,$\sigma_s$=5,$d$=5   & 22.54   & 0.5872   & \textbf{23.40}   & 0.6038   & \textbf{23.40}   & \textbf{0.6149} \\ \hline 
$\sigma_g$=50,$\sigma_s$=5,$d$=25   & 19.71   & 0.5162   & 22.55   & 0.5471   & \textbf{22.67}   & \textbf{0.5786} \\ \hline 
$\sigma_g$=50,$\sigma_s$=25,$d$=5   & \textbf{20.59}  & 0.5390   & 20.49   & 0.5063   & \textbf{20.59}   & \textbf{0.5635} \\ \hline 
$\sigma_g$=50,$\sigma_s$=25,$d$=25   & 19.22   & 0.4850   & 21.46   & 0.4869   & \textbf{21.86}   & \textbf{0.5673} \\ \hline 
\end{tabular}
\end{table*}

\subsection{Denosing Results on Fusion Noise (Unknown Noise Statistics)}

\label{sec:blind_noise}

In this section, we compare the performance of denoising  methods  when  the  prior  knowledge  of  noise  statistics  is not  available.  
For this, we generated the fusion noise, which is a mixture of different noise models. 
We combined three known noise models, AWGN, speckle noise, and salt-and-pepper noise, with $\sigma_g$, $\sigma_s$, and $d$ to simulate this fusion noise, which is formally defined as follows: 
%
\begin{gather}
y = f_{spn}((x+n_g)+n_s*(x+n_g),d) \label{eq:fusion}\\ 
n_g \sim \mathcal{N}(0,\,\sigma^{2}_{g})\,,
n_s \sim \mathcal{U}(0,\,\sigma^{2}_{s})\,.
\end{gather}
To compare the results on various noise levels, we selected $\sigma_{g}\in \{25,50\}$, $\sigma_{s}\in \{5,25\}$, and $d\in \{5,25\}$.

%
%
%
%
We compare N2K with three well-known blind denoisers (N2V, N2S, and S2S), along with N2C (supervised denoiser) as a baseline. 
%
For highly corrupted images, the image prior knowledge related to gradient can improve the denoising performance.
Hence, in order to achieve the best performance, we employed a TV regularization term in N2K as shown below: 
%

\begin{gather}
    \mathcal{L}(f,w) = \frac{1}{N}\sum_{i}^{N}(\sum_{J}w^{i}_{J}(f_J(x^{i})-x^i_J)^2 + \alpha ||f(x^i)||_{TV})
\end{gather}

As shown in Fig.~\ref{fig:qualitative_for_fusion}, all other blind denoising methods inaccurately reconstructed the a black color to brighter gray color. 
%
%
%
N2S and S2S also suffered from structural artifacts as well as incorrect brightness (Fig.,~\ref{fig:qualitative_for_fusion} last row).
%
We observed that N2K predicted the clean image more accurately while preserving the image contrast and details well as compared to N2S, N2V, and S2S. 
%
Furthermore, the TV-added version of N2K, called N2K+TV, effectively removed noise while preserving sharp edges. 

Table~\ref{table:fusion} summarizes the results for various noise levels and denoising methods; N2K and N2K+TV achieved the highest PSNR compared to the state-of-the-art blind denoising methods. 
It is clearly shown that the TV regularization effectively improves SSIM, especially for the higher noise levels. 
%
%
%
We also observed that the performance gap between N2K and the others becomes larger as the noise level increases. 
%
In summary, we conclude that N2K and N2K+TV overcome the problems caused by the non-zero mean noise that affects most other denoising methods.

\subsection{Ablation Study}

In this section, we empirically show the difference in the performance of ADSS loss against the general self-supervision loss. 
In this experiment, we used the same network structure for all test cases; however, the network was trained using different loss functions to see how they affected the performance. 
The baseline model was trained using the general self-supervision $L_2$ loss (\ref{eq:selfloss}), N2K is trained using the proposed ADSS loss, and N2K+TV is trained using the ADSS loss with a total variation term. 
%
%
%
%
%
%
%

Table~\ref{table:ablation_adss} shows the results of the previously introduced models when tested on the BSD68 dataset.  
%
%
It can be seen that the ADSS loss, which suppresses training from outliers, 
outperforms the general self-supervision loss at various levels of fusion noise except the case of $\sigma_g$=50, $\sigma_s$=5, and $d$=5. 
In addition, ADSS+TV achieved higher PSNR and SSIM than the baseline and ADSS alone.
Since the general self-supervision loss considers all pixels to be training data, it is more sensitive to highly corrupted noisy pixels. 
The study result also confirms that the performance gap between baseline and N2K is bigger for higher noise levels ($d = 25$). 
%
As shown in the unknown noise statistics experiment, we observed that the TV loss helped to increase PSNR and SSIM in highly corrupted images. 
%
%
%

In this ablation study, we demonstrated that the ADSS loss outperformed the general self-supervision loss. 
We also observed that the TV regularization was highly effective at further improving the image quality.
%

\section{Conclusion}
In this paper, we introduced a novel unsupervised denoising method, \texttt{Noise2Kernel}.
The proposed method is based on the dilated convolutional $\mathcal{J}$-$invariant$ network, allowing for efficient kernel-based training without the masking scheme.
The absence of preprocessing further pushes the performance in terms of training efficiency.
We also proposed an adaptive self-supervision loss that is highly effective in preserving overall brightness and structures in the image, even with the extremely high noise level and even if the zero mean assumption and prior knowledge of noise statistics are not present.
Using simulations of known and unknown noise statistics, we showed that N2K leads to better denoising quality than other state-of-the-art methods of blind denoising.
We believe the proposed work will be useful in improving highly corrupted noisy images where noise statistics are not readily available. 
In the future, we plan to explore applications of our method, especially in the medical imaging domain. 
Extending the proposed architecture to general image enhancement problems, such as blind image super-resolution, is another interesting future work.

\bibliographystyle{unsrt}  
\bibliography{egbib}

\end{document}


\floatsetup[subfigure]{capbesideposition={left,top}}
\theoremstyle{plain}
\newtheorem{thm}{Theorem} 

\theoremstyle{definition}
\newtheorem{defn}{Definition} 
\newtheorem{prop}[thm]{Proposition}

\pagestyle{headings}
\mainmatter
\def\ECCVSubNumber{6234}  
\newcommand*{\QEDA}{\hfill\ensuremath{\blacksquare}}
\title{Noise2Kernel: Blind Denoising with End-to-End Adaptive Self-Supervised Learning} 

\titlerunning{ECCV-20 submission ID \ECCVSubNumber} 
\authorrunning{ECCV-20 submission ID \ECCVSubNumber} 
\author{Anonymous ECCV submission}
\institute{Paper ID \ECCVSubNumber}

\maketitle
In this supplementary note, we provide the complete proof of Proposition 2 and additional qualitative results. 
We pick salt-and-pepper noise and fusion noise because these examples demonstrate a clear benefit of N2K comparing to other state-of-the-art methods (i.e., N2V~\cite{noise2void} and N2S~\cite{noise2self}).
%

\section{Supplementary Proof of Proposition 2}

Let $y \in \mathbb{R}^{m_1\times m_2}$ be arbitrary random variables and $\mathcal{I}$ be a set of dimensions $\{1,...,m_1\}\\
\times\{1,...,m_2\}$. 
%
%
We set the function $f$ as the network defined in the paper. 
We demonstrate that $f$ is an invariant function if $d \geq \ceil{K/2}$. 
Due to the donut convolution layer, the $f^{(0)}_{i,j}$ information moves to the follow dimensions:
\begin{gather}
\nonumber
\{i- \floor{K/2},...,i-1,i,i+1,...,i+\floor{K/2}\}\times \\
\{j- \floor{K/2},...,j-1,j,j+1,...,j+\floor{K/2}\} \setminus \{i\}\times\{j\} \subset \mathcal{I}. 
\label{eq:1}
\end{gather}
To exclude the information $y^{(0)}_{i,j}$ at $f^{(1)}(f^{(0)}(y^{(0)}))_{i,j}$, $d$ of $f^{(1)}$ should be greater than or equal to $\ceil{K/2}$ obviously.

We only consider the cases of $d \geq \ceil{K/2}$. 
Assume $n$ is the number of $d$-dilated convolution layers and $k \in [2,n]$. 
Then $y^{(k)}_{i,j}$ is generated by $f^{(k-1)}_{i,j}$ $d$-dilated convolution layer applied to $y^{(k-1)}$ as follows:
%
%
\begin{gather}
y^{(k)}_{i,j} = \Sigma_{p=q=1}^{3}w_{p,q}*y^{(k-1)}_{i+(p-2)*d,j+(q-2)*d}
\label{eq:2}
\end{gather}
and the information from $y^{(k-1)}_{i,j}$ moves to the following dimensions: 
\begin{gather}
\{i-d, i, i+d\} \times \{j-d,j,j+d\} \subset \mathcal{I}. 
\label{eq:3}
\end{gather}
%
Similarly, we can define the region $\mathcal{S}_2$ , which is independent from $y^{(0)}_{i,j}$,  iteratively from $y^{(1)}$ to $y^{(n)}$ as follows: 
%
%
\begin{gather}
\nonumber
\{i\} \times \{j\} \rightarrow \\
\nonumber
\{i-d, i, i+d\} \times \{j-d, j, j+d\} \rightarrow  \\
\nonumber
\{i-2d,i-d,i,i+d,i+2d\} \times \{j-2d,j-d,j,j+d,j+2d\} \rightarrow  \\
\nonumber
... \rightarrow \\
\nonumber
\{i-nd,i-(n-1)d,...,i-d,i,i+d,...,i+(n-1)d,i+nd\} \times\\
\{j-nd,j-(n-1)d,...,j-d,j,j+d,...,j+(n-1)d,j+nd\}
\label{eq:4}
\end{gather}
As shown above, the information from $y^{(0)}_{i,j}$ never reaches $f(y^{(0)})_{i,j}$ if $d \geq \ceil{K/2}$.\QEDA

\section{Additional Qualitative Results}
\begin{figure}[h]
\centering
\includegraphics[height=13.3cm,keepaspectratio]{eccv2020kit/SP_50_supplement.pdf}
   \caption{Qualitative performance comparison of various denoising methods on salt-and-pepper noise model where $d$=50. The conventional indicates AT2F~\cite{singh2018adaptive} method.}
\end{figure}

\begin{figure}[H]
\includegraphics[height=12.5cm,keepaspectratio]{eccv2020kit/Fusion_25_25_25_supplement.pdf}
   \caption{Qualitative performance comparison of salt methods on unknown (i.e., fusion) noise model where $\sigma_g$=25, $\sigma_s$=25, and $d$=25. Three representative images are shown here.}
\end{figure}

\begin{figure}[H]
\includegraphics[height=12.8cm,keepaspectratio]{eccv2020kit/Fusion_50_25_25_supplement.pdf}
   \caption{Qualitative performance comparison of various denoising methods on unknown (i.e., fusion) noise model where $\sigma_g$=50, $\sigma_s$=25, and $d$=25. Three representative images are shown here.}
\end{figure}

\FloatBarrier

\bibliographystyle{splncs04}
\bibliography{egbib}